# Securing Enterprise Email Against Quantum Threats: A Ten-Layer Hybrid Architecture and Its Operational Costs

Akshaya C[1] Prathiksha R[1] Vasanthi B[1] Aswani Kumar Cherukuri[1,2]

[1]School of Computer Science Engineering and Information Systems

[2]Centre For Artificial Intelligence Research (C-FAIR)

Vellore Institute of Technology, Vellore 632014, India.

Email: cherukuri@acm.org

## Abstract

Email is protected today by RSA, elliptic-curve and Diffie-Hellman cryptography, all of which a large quantum computer could break. Encrypted mail that is captured now can be stored and decrypted later (harvest now, decrypt later), and signatures can be forged once the classical keys are broken. This work presents a hybrid post-quantum secure email architecture that adds the NIST-standardised ML-KEM-768 (FIPS 203) and ML-DSA-65 (FIPS 204) algorithms to every layer that uses public-key cryptography, while keeping classical algorithms alongside them. The design covers end-to-end S/MIME (X25519 + ML-KEM-768 key establishment through HKDF, AES-256-GCM encryption and ECDSA + ML-DSA dual signatures), hybrid DKIM (RSA + ML-DSA), SPF and DMARC evaluation, TLS 1.3 with DANE/TLSA authentication that includes an ML-DSA proof, and POP3 over TLS delivery. Verification requires both the classical and the post-quantum signature, which prevents downgrade attacks, and every received message is given a protection level so that weaker or unrecognised protection is never treated as safe. A working prototype was built in Python and evaluated on effectiveness, security, reliability and efficiency. All 136 automated tests pass, the full suite passed in 10 consecutive runs, 100 randomised end-to-end cycles succeeded, and the ML-KEM-768 and ML-DSA-65 implementations match 100 official NIST test vectors. Compared with a classical-only baseline, the hybrid design adds a constant 7,336 bytes per encrypted message, about 5.4 KB per email for the two DKIM headers and a 2.6 KB DKIM DNS record that is likely to need DNS over TCP. Against a simulated quantum attacker, the classical-only baseline failed all three attack scenarios while the hybrid design resisted all three. End-to-end delivery on a single machine averaged 187.6 ms with a pure-Python post-quantum backend. The results show that a hybrid post-quantum email system is feasible, and that its main cost is message and DNS size rather than computation.

## 1. Introduction

Email is one of the most widely used communication systems and it depends on public-key cryptography at several independent layers. Message encryption and signing (S/MIME) use RSA or elliptic curves, domain authentication (DKIM) uses RSA signatures, transport security (TLS) uses elliptic-curve key exchange, and server authentication through DANE publishes certificate hashes in DNS. The mathematical problems behind RSA, elliptic-curve and Diffie-Hellman

cryptography can be solved efficiently by Shor's algorithm on a large quantum computer [17], so all of these layers would be exposed together.

The threat is not only in the future. An adversary who records encrypted mail today can decrypt it once such a computer exists. Email is a particularly exposed case because messages are stored for years. Signatures face a related risk: once a classical signing key can be derived from its public key, forged messages can be made to look genuine. For this reason, the move to post-quantum cryptography cannot wait until a quantum computer appears.

The National Institute of Standards and Technology (NIST) has standardised two post-quantum algorithms: ML-KEM for key establishment [18] and ML-DSA for digital signatures [19]. However, these algorithms are new and have less field experience than RSA and elliptic curves. A practical migration therefore keeps the classical algorithms and adds the post-quantum ones beside them (a hybrid design), so that the system stays secure if either family is broken. The remaining questions for email are how to apply this across every layer, how to prevent an attacker from forcing a message back to classical-only protection, and what the change costs in message size and time.

This study is motivated by the following research questions: Can a hybrid post-quantum design be applied to every public-key layer of email in a working system? Does it resist attacks, including a quantum-capable attacker, where classical-only protection fails? What does it cost compared with the classical design?

The main objectives of this project are as follows. First, to design a hybrid architecture that adds ML-KEM-768 and ML-DSA-65 to S/MIME, DKIM, DANE and the mail delivery path while keeping classical algorithms. Second, to build a working prototype that sends, stores, downloads, decrypts and verifies real messages. Third, to evaluate the prototype on effectiveness, security, reliability and efficiency using automated tests, randomised runs and an independent check against NIST test vectors. Finally, to compare the hybrid design with a classical-only baseline on the same pipeline, including a simulated quantum attacker.

The project is limited to a software prototype running on a single machine with simulated DNS security (DNSSEC), and it does not include a production deployment. This report is organised into nine sections. Section 2 presents the literature survey. Section 3 describes the problem formulation. Section 4 explains the proposed method and architecture. Section 5 describes the experimental setup. Section 6 presents the results and comparative analysis. Section 7 discusses the findings and limitations. Section 8 concludes the study and outlines future work, and Section 9 lists the references.

## 2. Literature Survey

This review surveys Post-Quantum Cryptography (PQC), covering its main algorithm families, implementation approaches, security concerns, performance evaluations, and real-world applications. The families examined include lattice-based, code-based, hash-based, multivariate, and isogeny-based cryptography, along with post-quantum key exchange, digital signatures, hybrid encryption, and quantum-safe communication. Common obstacles to adopting PQC include computational overhead, memory demands, large key sizes, hardware constraints, and implementation-level attacks.

One group of studies focuses on migration and security analysis of existing systems. One work uses large language models such as GPT-4.1 and CodeLlama to convert classical cryptographic code into post-quantum code using ML-KEM and ML-DSA, though migrating large projects with multiple files and dependencies remains difficult [1]. Another introduces a framework to verify whether S/MIME emails are protected by post-quantum cryptography, noting that insecure classical paths based on RSA and ECDH may still persist [2]. A further study examines the usability of S/MIME certificates containing multiple email addresses and finds that existing email clients offer limited support for them [11]. A roadmap-based study analyses the threats posed by Shor's and Grover's algorithms and outlines a transition to post-quantum security, but the roadmap needs better adaptability across organizations [15].

A second group reviews and compares algorithms. One implementation study covers lattice-based schemes such as LWE, Ring-LWE, Kyber, Frodo, and NewHope for secure communication, but requires more testing for performance, scalability, and resource-limited devices [3]. Another compares lattice-based, code-based, hash-based, and isogeny-based algorithms by speed, memory, and key size, and calls for more efficient implementations and migration strategies [4]. A comprehensive review of post-quantum digital signatures covers lattice-based, code-based, hash-based, and multivariate schemes, highlighting the need for efficient and secure signatures suited to real-world use [5].

A third group proposes new schemes and optimizations. A hybrid framework combines Kyber with AES and adds real-time detection of thermal and electromagnetic attacks, though it needs testing on varied hardware platforms [6]. A new key exchange protocol based on conic intersections, algebraic geometry, and Veronese embedding still requires further security analysis and performance testing [7]. A hardware processor for NTT/INTT with Montgomery reduction improves FALCON key generation, but it addresses only part of the complete FALCON system [8]. A code-based approach using the Niederreiter cryptosystem, Fujisaki-Okamoto transform, and sparse matrices reduces key size, processing time, and energy consumption, with scalability and side-channel security still to be tested [9]. An AVX2-optimized implementation of the CROSS signature scheme, using SHAKE and the Fiat-Shamir transform, improves efficiency but calls for further performance gains and practical implementation work [16].

A final group addresses hybrid and application-specific deployments. One framework secures legacy Modbus industrial control systems using Kyber, Dilithium, decentralized identifiers, and verifiable credentials, though it depends on proxy deployment and better device management [10]. A hybrid TLS design for software-defined networks combines classical key exchange, ML-KEM, ML-DSA, and Quantum Key Distribution (QKD), and needs evaluation in large, complex networks [12]. Another work proposes secure group communication using Kyber, BIKE, HQC, Falcon, Dilithium, and QKD, with group authentication and scalability needing improvement [13]. A communication framework for industrial federated learning built on ML-DSA, ML-KEM, SHA3-512, and AES-256-GCM requires more testing in real industrial environments [14]. A related two-plane bridge architecture binds a TLS 1.3 exporter secret with an application-layer ML-KEM-768 exchange authenticated via a pinned ML-DSA-65 key without modifying TLS libraries, though it targets point-to-point transport channels rather than multi-layer email protocols [39].

Taken together, these studies show steady progress in post-quantum techniques for encryption, key exchange, digital signatures, secure communication, and industrial settings. Significant challenges persist, however, including scalability, limited hardware support, side-channel vulnerabilities, complex migration of existing systems, and a lack of real-world testing. Further research is needed to deliver PQC solutions that are efficient, scalable, and secure enough for practical deployment.

The table 1, categorises existing literature spanning LLM-assisted PQC code migration, S/MIME security analysis, lattice- and code-based optimizations, hardware accelerators, and hybrid deployments in SDN and industrial control systems, highlighting persistent challenges in scalability, side-channel resistance, and practical end-to-end integration.

Table 1: Comparative Summary of Post-Quantum Cryptography Literature (2014–2026)

| S.No. | Paper Title (Year) | Algorithms Used | Proposed Work / Methodology | Research Gap |
|---|---|---|---|---|
| 1 | Empirical Evaluation of Large Language Models for Migration of Code Fragments to PQC (2026) | GPT-4.1, CodeLlama, ML-KEM, ML-DSA | Uses AI models to convert traditional cryptographic code into post-quantum cryptography code. | Difficult to migrate large projects with multiple files and dependencies. |
| 2 | Analysing the Post-Quantum Security of S/MIME (2026) | ML-KEM, ML-DSA, RSA, ECDH | Introduces a framework to check whether S/MIME emails are protected using post-quantum cryptography. | Existing systems may still contain insecure classical encryption paths. |
| 3 | Implementing Lattice-Based Cryptosystems for PQC (2026) | LWE, Ring-LWE, Kyber, Frodo, NewHope | Implements different lattice-based cryptographic algorithms for secure communication. | More testing is needed for performance, scalability, and resource-limited devices. |
| 4 | Post-Quantum Cryptography Algorithm's Standardization and Performance Analysis (2022) | Lattice-based, Code-based, Hash-based, Isogeny-based | Compares different post-quantum algorithms based on speed, memory, and key size. | More efficient implementations and migration strategies are required. |
| 5 | Securing the Future: A Comprehensive | Lattice-based, Code-based, | Reviews different post-quantum digital signature | Need for efficient and secure signature |
| | Review of Post-Quantum Digital Signatures (2026) | Hash-based, Multivariate | schemes and their performance. | schemes for real-world applications. |
| 6 | PSESV: A Hybrid Post-Quantum Encryption Framework (2025) | Kyber, AES, TEHM, DDCV, RCRR | Combines post-quantum encryption with real-time detection of thermal and electromagnetic attacks. | Requires further testing on different real-world hardware platforms. |
| 7 | A Post-Quantum Key Exchange Protocol from the Intersection of Conics (2025) | Algebraic Geometry, Conic Intersection, Veronese Embedding | Proposes a new key exchange method based on mathematical properties of conics. | Needs further security analysis and performance testing. |
| 8 | Efficient NTT/INTT Processor for FALCON Post-Quantum Cryptography (2025) | FALCON, NTT, INTT, Montgomery Reduction | Develops an efficient hardware processor to improve FALCON key generation. | Focuses mainly on NTT/INTT rather than the complete FALCON system. |
| 9 | Optimizing Code-Based Cryptography for Efficient and Secure PQC Key Agreement (2025) | Niederreiter, Fujisaki-Okamoto, Sparse Matrices | Improves code-based cryptography by reducing key size, processing time, and energy consumption. | Further testing is needed for scalability and side-channel security. |
| 10 | Securing Modbus in Legacy Industrial Control Systems (2025) | Kyber, Dilithium, DIDs, Verifiable Credentials | Uses post-quantum cryptography and decentralized identity to secure legacy industrial systems. | Depends on proxy deployment and requires further improvements in device management. |
| 11 | S/MIME with Multiple E-mail Address Certificates: A Usability Study (2014) | S/MIME, X.509, RSA, DSA, AES | Studies the usability of certificates containing multiple email addresses. | Existing email clients have limited support for multiple-address certificates. |
| 12 | Hybrid Quantum-Safe Integration of TLS in SDN Networks (2025) | ML-KEM, ML-DSA, DHKE, QKD | Combines classical cryptography, post-quantum cryptography, and quantum key distribution in TLS. | Needs further evaluation in large and complex network environments. |
| 13 | Hybrid Quantum-Secure Group-Based Communication with PQC and QKD (2025) | Kyber, BIKE, HQC, Falcon, Dilithium, QKD | Proposes secure group communication methods using PQC and QKD. | Group authentication and scalability need further improvement. |
| 14 | PQC-FLCom: Post-Quantum Secure Communication for Industrial Federated Learning (2026) | ML-DSA, ML-KEM, SHA3-512, AES-256-GCM | Develops a secure communication framework for industrial federated learning. | More practical testing is needed in real industrial environments. |
| 15 | Complexity and the Transition to Post-Quantum Security (2025) | Shor's Algorithm, Grover's Algorithm, Kyber, AES-256 | Studies quantum computing threats and provides a roadmap for transitioning to post-quantum security. | The proposed roadmap needs better adaptability to different organizations. |
| 16 | A High-Efficiency AVX2-Optimized Engineering of the Post-Quantum Digital Signature CROSS (2026) | CROSS, SHAKE, AVX2, Fiat-Shamir | Improves the performance of the CROSS digital signature scheme using AVX2 optimization. | Further improvements in performance and practical implementation are needed. |
| 17 | Hybrid Two-Plane Approach for Post-Quantum Cryptographic Migration in Transport Layer Security, 2026 | TLS 1.3 (X25519), ML-KEM-768, ML-DSA-65, HKDF-SHA-256, AES-256-GCM | Proposes a library-agnostic two-plane bridge that binds a TLS 1.3 exporter secret with an application-layer ML-KEM-768 exchange authenticated by a | Focuses on point-to-point TLS channels and requires application-level framing and out-of-band key pinning, leaving multi-layer store-and-forward email protocols (S/MIME, DKIM, |
| | | | pinned ML-DSA-65 server key without modifying existing TLS libraries. | DMARC, DANE) unaddressed. |

## 3. Problem Formulation

An email system consists of a sender, a sending server, a receiving server, a mailbox server and a receiving client. In the classical system, public-key cryptography is used at ten layers L1 to L10 (key management, message encryption, message signature, DKIM, SPF, DMARC, TLS, DANE, DNS security and mailbox delivery). The adversary is assumed to have the following capabilities:

- **A1:** observe and record all network traffic and store it indefinitely;
- **A2:** modify, replay, drop or inject messages and DNS data in transit;
- **A3:** run their own mail server, domain and keys, and claim to be another user;
- **A4:** at a later time, recover classical private keys from public keys using a quantum computer, but not break ML-KEM-768 or ML-DSA-65;
- **A5:** attempt to abuse services through open relay, password guessing, header injection, path traversal or oversize messages.

The adversary is not assumed to compromise the endpoints or to obtain private keys by other means. Table 2 lists the security and practicality goals that the system must satisfy, and each goal is linked to the evaluation tests in Section 5.

**Table 2: Security and Practicality Goals**

| Goal | Requirement | Adversary |
|---|---|---|
| Confidentiality | Mail recorded today stays secret even if the classical keys are later broken | A1, A4 |
| Integrity | Any change to the body, headers, ciphertext or key data is detected | A2 |
| Authenticity | A message from a forged sender is rejected at message level and domain level | A3, A4 |
| Server authenticity | A fake server is rejected before any mail is sent | A2, A3 |
| Downgrade resistance | Removing or invalidating post-quantum protection never lowers a message to classical-only trust | A2 |
| Abuse resistance | Relay, brute-force, injection, traversal and oversize attempts are refused | A5 |
| Practicality | The extra size and time are measurable and acceptable | - |

Let $ss_x$ be the X25519 shared secret and $ss_k$ the ML-KEM shared secret. The message key is derived from both secrets together with the one-time public key and the KEM ciphertext, so the key stays secret if either secret remains secret:

$$\text{CEK} = \text{HKDF}(ss_x, ss_k, \text{eph_pub}, \text{kem_ct}) \tag{1}$$

For signatures, a message m with an ECDSA signature $s_c$ and an ML-DSA signature $s_p$ is accepted only if both checks succeed:

$$\mathrm{accept}(m) = \mathrm{Verify}_{\mathrm{ECDSA}}(m, s_c) \wedge \mathrm{Verify}_{\mathrm{MLDSA}}(m, s_p) \quad (2)$$

A message is therefore forgeable only if both the classical and the post-quantum signature can be forged. The evaluation is organised around five research questions: RQ1, does the system work correctly under normal use (effectiveness); RQ2, does it reject the attacks A1 to A5 (security); RQ3, does it behave consistently under repeated and randomised use (reliability); RQ4, what does hybrid cost compared with classical in size and time (efficiency); and RQ5, does hybrid hold against a quantum-capable attacker where classical-only fails.

## 4. Proposed Method and Architecture

Figure 1 shows the proposed architecture. A message is signed and encrypted by the sender's client, signed again at domain level by the sending server, sent over TLS 1.3 to a receiving server that is verified through DANE, checked with SPF, DKIM and DMARC, stored, and finally downloaded over TLS by the receiving client, which decrypts it and verifies the signatures. DNS publishes the DKIM keys, the SPF and DMARC policies and the TLSA records.

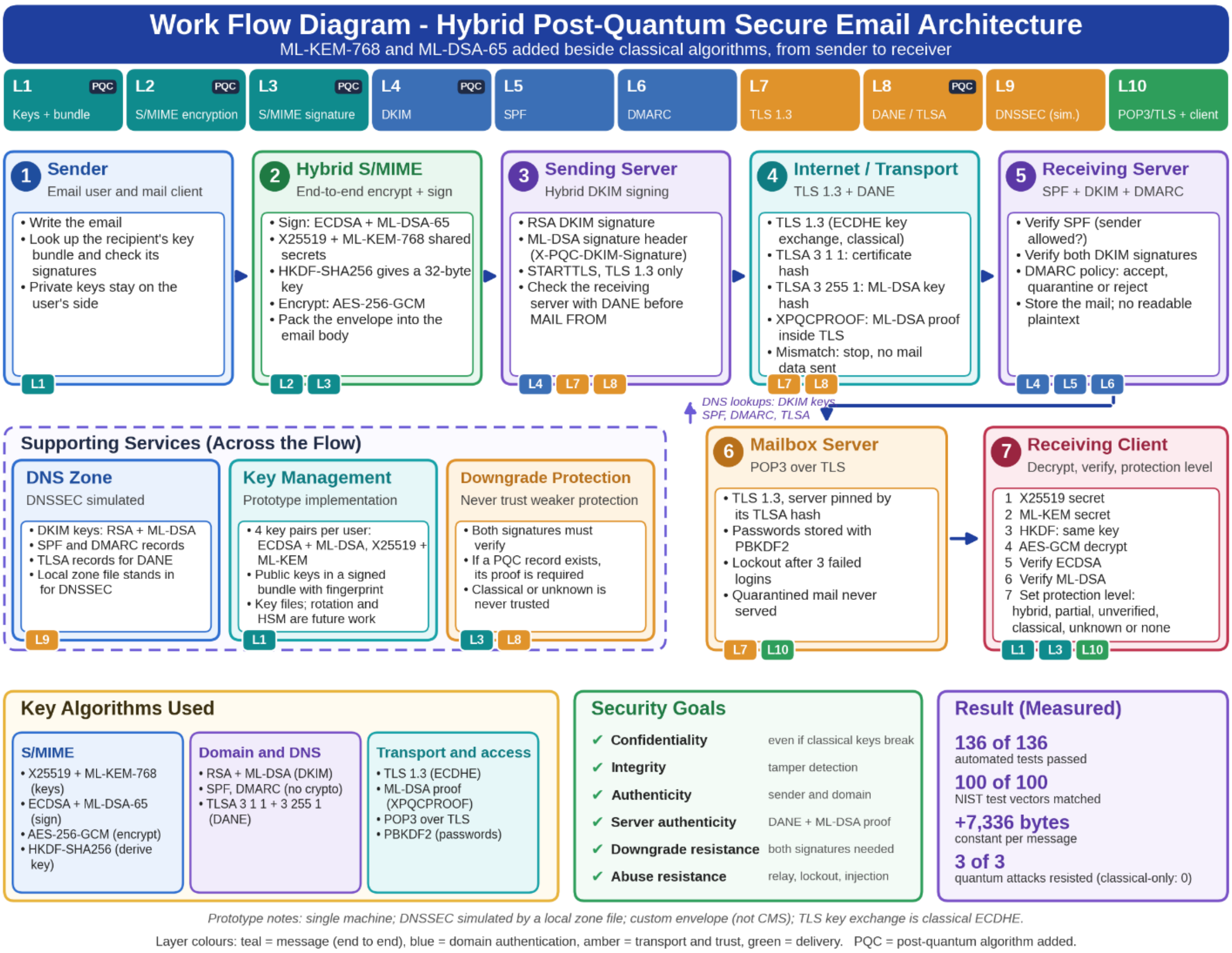

**Figure 1: Architecture of the Hybrid Post-Quantum Secure Email System**

### 4.1 Overview

The architecture has ten layers. Layers 1 to 3 protect the message end to end, layers 4 to 6 authenticate the sending domain, layers 7 to 9 establish trust in the servers, and layer 10 delivers the mail to the user. Post-quantum algorithms are added at the layers that use public-key signatures or key establishment, and classical algorithms are kept beside them. SPF uses no cryptography and needs no post-quantum change, TLS key exchange remains classical, and DNSSEC is simulated in this prototype. Description of each layer is provided in Table 3.

**Table 3: Security Layers and Post-Quantum Additions**

| Layer | Mechanism | Post-quantum addition |
|---|---|---|
| L1 | Key management and signed key bundles | ML-DSA bundle signature; ML-KEM key |
| L2 | S/MIME encryption | X25519 + ML-KEM-768, HKDF, AES-256-GCM |
| L3 | S/MIME signature | ECDSA + ML-DSA-65, both required |
| L4 | DKIM | RSA + ML-DSA signature header |
| L5 | SPF | None (no cryptography) |
| L6 | DMARC | Alignment with the hybrid DKIM result |
| L7 | TLS 1.3 (STARTTLS) | Classical ECDHE (not post-quantum) |
| L8 | DANE / TLSA | ML-DSA proof and binding to the certificate |
| L9 | DNSSEC | Simulated by a local zone file |
| L10 | POP3 over TLS and client | Server pinning; client verifies hybrid signatures |

### 4.2 Key Management

Each user holds four key pairs: ECDSA and ML-DSA for signing, and X25519 and ML-KEM for encryption. The public keys are published in a bundle that is signed by both signing keys and checked with a fingerprint, so a modified or swapped bundle is rejected. Private keys stay on the user's side, and the mail servers never load private keys or the S/MIME code.

### 4.3 End-to-End S/MIME Encryption and Signature

The sender signs the message first and then encrypts it, so the signatures are hidden inside the encrypted data [33]. Figure 2 shows the steps on both sides. The sender looks up the recipient's bundle, signs the message with ECDSA [20] and ML-DSA-65, generates a one-time X25519 key and derives a shared secret with the recipient's X25519 key [22], and encapsulates to the recipient's ML-KEM-768 [18,19] key, which gives a 1,088-byte ciphertext and a second shared secret. Both secrets are combined with HKDF-SHA256 [21] into a 32-byte key, the message and both signatures are encrypted with AES-256-GCM [23], and the one-time key, KEM ciphertext, nonce and ciphertext are packed into an envelope in the email body.

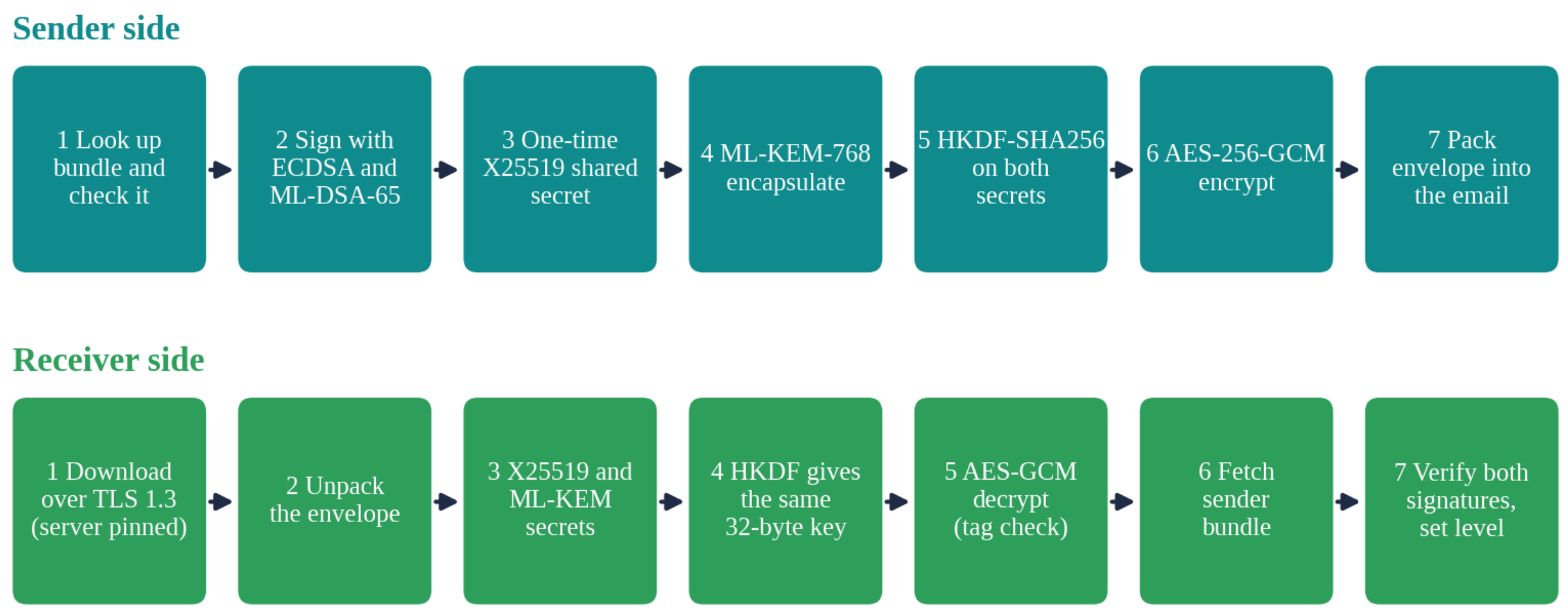

**Figure 2: S/MIME Encryption and Decryption Flow**

The receiver reverses these steps. It derives both secrets, rebuilds the same key, and decrypts the envelope, where any change makes the authentication tag fail. It then fetches the sender's bundle from the From address and verifies both signatures. The result is a status such as ok, decrypt_failed, bad_signature or unknown_sender.

### 4.4 Domain Authentication

Hybrid DKIM [24] adds a second signature header, X-PQC-DKIM-Signature, which uses ML-DSA beside the RSA DKIM signature. Both signatures must verify, and the two public keys are published in DNS. SPF [25] is evaluated by an own checker that supports ip4, ip6, include, redirect and the all qualifiers, and an unsupported mechanism gives a safe permerror. DMARC [26] requires the From domain to align with the SPF result or with the hybrid DKIM result, and applies the domain policy: reject (550 reply), quarantine (kept in a separate folder) or accept.

### 4.5 Server Trust with TLS 1.3 and DANE

The sending server uses STARTTLS and accepts only TLS 1.3 [27]. It then checks the receiving server using DANE [28,29]. The TLSA records contain the certificate hash (usage 3, selector 1, matching type 1) and a second record for the ML-DSA key, which uses selector 255 in this prototype. The receiving server proves possession of the ML-DSA key through a custom XPQCPROOF command inside the TLS session, and the key must be bound to the certificate. Any mismatch stops the sender before the MAIL FROM command, so no message data leaves. If a post-quantum record is published, removing it is treated as a downgrade and is rejected. DNSSEC is simulated by treating a local zone file as trusted.

### 4.6 Mail Delivery and Client

The POP3 server [30] runs over TLS 1.3. The client pins the server to its TLSA hash before it sends any password. Passwords are stored with PBKDF2, an account is locked after three failed logins, and quarantined mail is never served. The client downloads the mail, decrypts it, verifies both signatures and reports a status.

### 4.7 Protection Level

Following the idea that unknown or weaker protection must never be treated as safe [5], the client assigns a protection level to every received message after decryption and verification (Table 4). The client reads the key-establishment field in the envelope header. An envelope with both X25519 and ML-KEM parts is hybrid, an envelope without an ML-KEM part is classical, and an unreadable or unrecognised envelope is unknown. For a hybrid envelope the level also depends on the verification result. The level is added to each message as an extra field and does not change the existing status checks.

**Table 4: Protection Levels of Received Messages**

| Level | Meaning |
|---|---|
| hybrid | Hybrid encryption opened and both signatures verified |
| partial | Decrypted, but the signature check failed, so the sender is not trusted |
| unverified | Hybrid envelope that could not be decrypted or checked |
| classical | No ML-KEM component; not quantum-safe |
| unknown | Envelope unreadable or algorithm not recognised |
| none | Not a post-quantum S/MIME message |

### 4.8 Implementation

The prototype is written in Python. ML-KEM-768 and ML-DSA-65 come from the pure-Python libraries kyber-py and dilithium-py, which are wrapped in one module so the backend can be replaced by a compiled library. The classical primitives, AES-GCM and HKDF come from the cryptography library. DKIM, DNS and the servers use dkimpy, dnspython and aiosmtpd. The receiving server listens on port 2525 and the POP3 server on port 2110. Messages follow the standard email format and protocols [31,32].

## 5. Experimental Setup

### 5.1 Experimental Environment

The prototype is implemented in Python in a virtual environment on a single Windows computer. The sending server, receiving server, mailbox server and clients run on the same machine, and DNS data is served from a local zone file. The libraries and settings are listed in Table 5. All timings therefore exclude real network latency, and the absolute values reflect a pure-Python post-quantum backend.

**Table 5: Experimental Environment**

| Item | Value |
|---|---|
| Machine | Single Windows computer |
| Language | Python |
| Post-quantum backend | kyber-py (ML-KEM-768), dilithium-py (ML-DSA-65), pure Python |
| Other libraries | cryptography, dkimpy, dnspython, aiosmtpd, pytest |
| Network | Local; servers and clients on one machine; DNS from a local zone file |
| Ports | Receiving server 2525; POP3 over TLS 2110 |

### 5.2 Evaluation Criteria and Metrics

The system is evaluated on four criteria that follow the quality characteristics of ISO/IEC 25010 [37]: effectiveness (functional suitability), security, reliability and efficiency (performance efficiency). Each percentage is computed from counts and not chosen in advance. The pass rate and the size overhead are defined as

$$\text{Pass rate} = (\text{tests passed} / \text{tests run}) \times 100\% \quad (3)$$

$$\text{Overhead} = (S_{\text{hybrid}} - S_{\text{classical}}) / S_{\text{classical}} \times 100\% \quad (4)$$

where S is the protected message size in bytes. For trial-based rates with k successes in n independent trials, a 95% Wilson score [38] interval is reported. Table 6 lists the metrics.

**Table 6: Evaluation Criteria and Metrics**

| Criterion | Metric | Method |
|---|---|---|
| Effectiveness | Functional pass rate | 70 functional tests |
| Security | Attack rejection rate | 60 attack tests; random tampering; 3 quantum-attacker scenarios |
| Reliability | Run success rate | 10 full-suite runs; 100 randomised end-to-end cycles |
| Efficiency | Size overhead; time as mean, SD and p95 | Same envelope at 1 KB, 10 KB and 100 KB |
| Scalability (partial) | Throughput in messages per second | Sequential cycles on one machine |

### 5.3 Test Suite

The evaluation uses 136 automated tests written with the pytest framework. The tests were designed from the system design, the applicable standards and the threat model, and they are not taken from an external test suite. For each mechanism they check the normal behaviour and the attacks against it, which is why rule-heavy mechanisms such as SPF, DMARC and DANE have more tests than simple primitives. Every test is recorded in a per-test table that lists what it checks, the standard it is based on, the threat it covers, its severity and its result. The tests are based on the standards in Table 7 and are not a formal conformance validation against them.

**Table 7: Basis of the 136 Tests**

| Basis | Test files | Tests |
|---|---|---|
| FIPS 203 and 204 [18,19] | pqc, metrics (size test) | 4 |
| Primitives: X25519, HKDF, ECDSA, AES-GCM [20, 21, 22, 23] | hybrid_kem, hybrid_sig, aead | 12 |
| Own design and threat model | smime_roundtrip, certs | 12 |
| DKIM, SPF, DMARC (RFC 6376, 7208, 7489) [24,25,26] | dkim_hybrid, spf_dmarc, dnszone | 46 |
| DANE and TLS (RFC 6698, 7672, 8446) [27, 28, 29] | dane, tls_dane | 19 |
| Email format, SMTP, POP3 (RFC 5322, 5321, 1939) [30, 31, 32] | compose, mta_pipeline, pop3_mua | 38 |
| Measurement of the prototype | metrics (timing and size) | 5 |
| **Total** | | **136** |

The 60 attack tests cover seven threats: tampering (10 tests), sender forgery (10), downgrade or weakening of one algorithm family (11), impostor server (10), loss of confidentiality (8), service abuse (8) and key or identity substitution (3). By severity, 43 of the tests are critical, 17 high, 70 medium and 6 low.

### 5.4 Experimental Procedures

- **Reliability:** the full suite was run 10 times in a row. In addition, 100 randomised cycles each generated fresh keys and a random message of 0 to 100 KB, protected it, opened it and compared the result. In every cycle one random bit of a random envelope field was also flipped, and the tampered message had to be rejected.
- **Timing:** after 3 warm-up runs, protect and open were repeated for classical and hybrid at 1 KB, 10 KB and 100 KB, and the mean, standard deviation, median and 95th percentile were recorded.
- **Classical baseline:** the same envelope with X25519 key agreement and ECDSA only, with no ML-KEM and no ML-DSA.
- **Quantum attacker simulation:** the attacker is given the classical private keys, which a quantum computer would recover from the public keys, and never the ML-KEM or ML-DSA keys. Three scenarios were run: harvest now, decrypt later; forged sender; and downgrade by stripping the post-quantum signature.
- **NIST known-answer check:** official ACVP test vectors from NIST's ACVP-Server repositor, in the format defined by the ACVP specifications [36], were fed to the ML-KEM-768 and ML-DSA-65 libraries and the output was compared with NIST's expected answers. The vectors cover ML-KEM key generation, encapsulation and decapsulation, and ML-DSA key generation and signature verification.
- **Protection level check:** seven message types were passed to the client: a genuine message, a forged sender, a classical-only envelope, a damaged ciphertext, an unknown sender, an unreadable envelope and an ordinary email.

## 6. Result and Comparative Analysis

The proposed hybrid post-quantum email system was evaluated using the automated test suite, repeated and randomised runs, a comparison with a classical-only baseline, a simulated quantum attacker and NIST test vectors. All 136 tests passed. The following subsections report the results for each criterion.

### 6.1 Effectiveness and Security

Table 8 shows that all 70 functional tests, all 60 attack tests and all 6 measurement tests passed, giving 136 of 136.

**Table 8: Test Results by Criterion**

| Criterion | Tests | Passed | Pass rate |
|---|---|---|---|
| Effectiveness (functional) | 70 | 70 | 100% |
| Security (attack) | 60 | 60 | 100% |
| Efficiency (measurement) | 6 | 6 | 100% |
| Total | 136 | 136 | 100% |

In the live demonstration, the sending server verified the receiving server through hybrid DANE (the classical and the post-quantum proofs were both true) over TLS 1.3. SPF, both DKIM signatures and DMARC passed, the server stored no readable plaintext, and the client recovered the message with status ok. In the randomised cycles, all 100 tampered messages were rejected.

### 6.2 Reliability

Table 9 shows the reliability results. The full suite passed in all 10 consecutive runs (1,360 of 1,360 test executions), and the run time varied from 26 to 96 seconds with machine load while the results did not change. In the randomised test, all 100 cycles succeeded, which gives a success rate of 100% with a 95% confidence interval of 96.30% to 100%. A run of 1,000 cycles without a failure would raise the lower bound to 99.62%.

**Table 9: Reliability Results**

| Test | Trials | Successes | Rate | 95% interval |
|---|---|---|---|---|
| Full suite, repeated runs | 10 | 10 | 100% | 72.2% to 100% |
| Randomised end-to-end cycles | 100 | 100 | 100% | 96.30% to 100% |
| Random tampering rejected | 100 | 100 | 100% | 96.30% to 100% |

### 6.3 Efficiency: Classical versus Hybrid

Table 10 and Figure 3 compare the protected message size of the classical baseline and the hybrid design. The hybrid design adds a constant 7,336 bytes at every message size. The relative overhead therefore falls from 340.3% for a 1 KB message to 4.1% for a 100 KB message, so small messages pay the most. Besides the envelope, the two DKIM headers add about 5.4 KB per email and the DKIM DNS record is about 2.6 KB, which is likely to need DNS over TCP.

**Table 10: Protected Message Size, Classical versus Hybrid**

| Plaintext | Classical | Hybrid | Extra bytes | Size overhead |
|---|---|---|---|---|
| 1,000 B | 2,156 B | 9,492 B | 7,336 B | 340.3% |
| 10,000 B | 18,156 B | 25,492 B | 7,336 B | 40.4% |
| 100,000 B | 178,156 B | 185,492 B | 7,336 B | 4.1% |

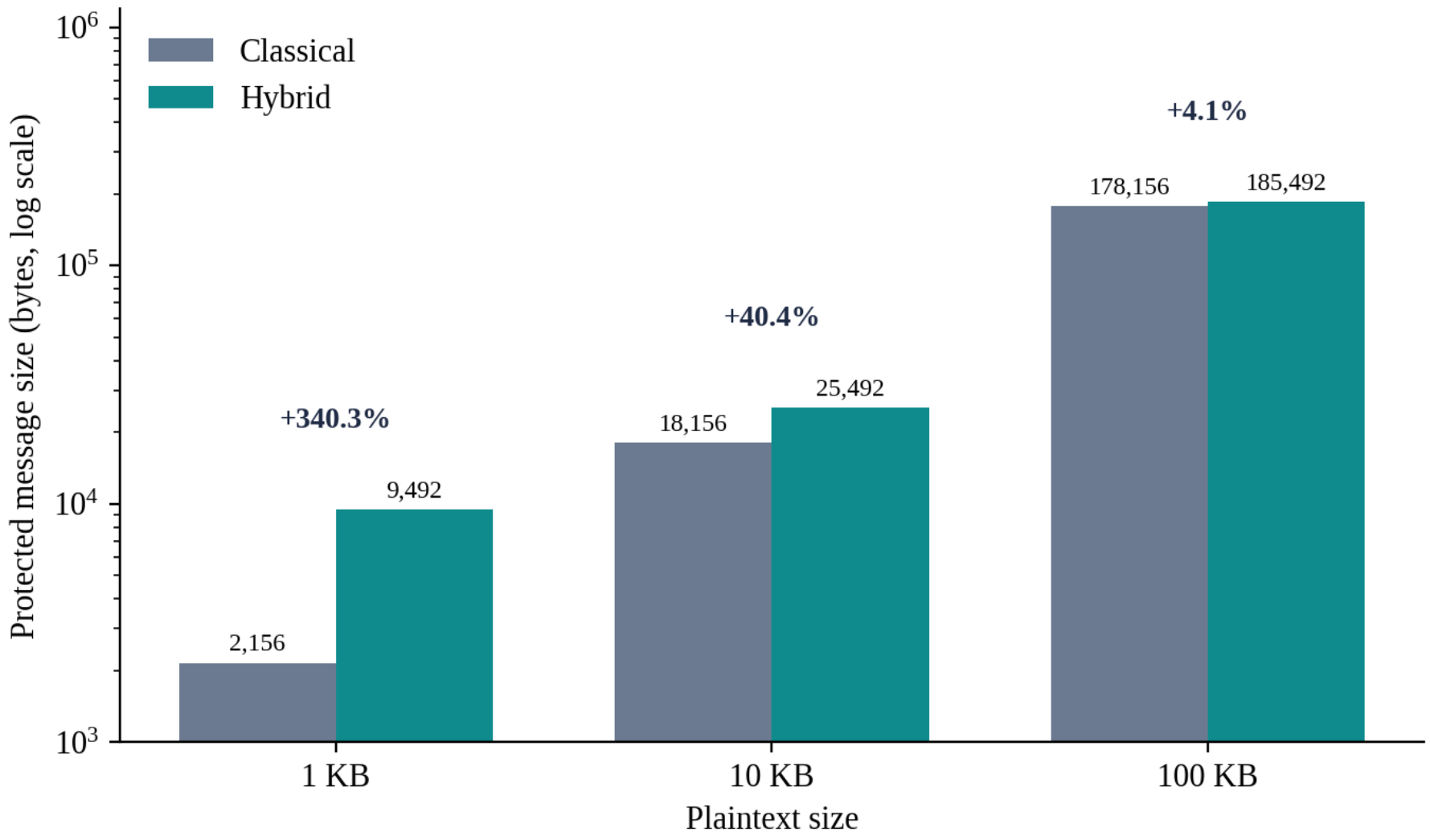

**Figure 3: Protected Message Size of the Classical and Hybrid Designs**

Table 11 give the time per operation. Opening a hybrid message (decrypting and verifying both signatures) is stable at about 15 ms. Protecting a hybrid message (signing with both algorithms and encrypting) takes 50 to 60 ms on average but varies widely, because ML-DSA signing repeats a random number of attempts until one is accepted, so the mean should be read together with the 95th percentile. The classical baseline runs in compiled code, whereas the post-quantum operations run in pure Python, so most of the ratio reflects the implementation language and not the algorithms. Sequential throughput for protect plus open was 13.6 messages per second, and end-to-end delivery averaged 187.6 ms.

**Table 11: Time per Operation in Milliseconds (Mean ± Standard Deviation, 10 Repeats)**

| Plaintext | Protect, classical | Protect, hybrid | Open, classical | Open, hybrid |
|---|---|---|---|---|
| 1,000 B | 0.11 ± 0.01 | 50.93 ± 29.14 (p95 122.95) | 0.11 ± 0.00 | 14.72 ± 0.49 |
| 10,000 B | 0.21 ± 0.00 | 60.15 ± 39.45 (p95 159.22) | 0.17 ± 0.00 | 14.46 ± 0.33 |
| 100,000 B | 1.76 ± 0.25 | 54.28 ± 24.44 (p95 108.42) | 1.01 ± 0.36 | 15.43 ± 0.68 |

### 6.4 Security: Classical versus Hybrid under a Quantum Attacker

Table 12 compares the two designs against the simulated quantum attacker. The classical-only design resisted none of the three scenarios (0 of 3) and the hybrid design resisted all three (3 of 3). Harvest now, decrypt later fails against the hybrid design because the message key also depends on the ML-KEM secret, which the attacker does not have. A forged sender fails because the attacker cannot produce a valid ML-DSA signature, and a downgrade fails because verification requires both signatures.

**Table 12: Classical-Only versus Hybrid under a Simulated Quantum Attacker**

| Attack | Classical only | Hybrid |
|---|---|---|
| Harvest now, decrypt later | Message decrypted | Stays secret |
| Forged sender (ECDSA broken) | Forgery accepted | Forgery rejected |
| Downgrade (post-quantum signature stripped) | Message accepted | Message rejected |
| **Scenarios resisted** | **0 of 3** | **3 of 3** |

### 6.5 Independent Check against NIST Test Vectors

To avoid testing the post-quantum libraries only with the project's own tests, they were run against NIST's official ACVP vectors (Table 13). All 100 vectors that were run passed and none failed. The 20 skipped ML-KEM vectors are key-validity checks and the 45 skipped ML-DSA vectors use the internal and pre-hash signature interfaces, which the project does not use. Signature generation vectors were not run. This is a check against NIST's published vectors and is not a formal validation.

**Table 13: Results of the NIST Known-Answer Check**

| Suite | Passed | Failed | Skipped |
|---|---|---|---|
| ML-KEM-768 key generation | 25 | 0 | 0 |
| ML-KEM-768 encapsulation and decapsulation | 35 | 0 | 20 |
| ML-DSA-65 key generation | 25 | 0 | 0 |
| ML-DSA-65 signature verification | 15 | 0 | 45 |
| **Total** | **100** | **0** | **65** |

### 6.6 Protection Level Results

Table 14 shows the protection level assigned in the seven-message check. Every message received the expected level, and neither the classical-only envelope nor the unreadable envelope was treated as safe.

**Table 14: Protection Level Assigned to Each Test Message**

| Message | Status | Protection level |
|---|---|---|
| Genuine hybrid message | ok | hybrid |
| Forged sender (signed by another user) | bad_signature | partial |
| Classical-only envelope (no ML-KEM) | malformed | classical |
| Ciphertext damaged in transit | decrypt_failed | unverified |
| Sender not in the key directory | unknown_sender | unverified |
| Unreadable envelope | malformed | unknown |
| Ordinary email without S/MIME | not_pqc | none |

### 6.7 Summary of the Comparison

Table 15 consolidates the paper's core empirical findings into a high-level side-by-side comparison between the traditional classical email baseline and the proposed 10-layer hybrid post-quantum architecture.

**Table 15: Summary of Classical-Only versus Hybrid Design**

| Aspect | Classical only | Hybrid |
|---|---|---|
| Message overhead | Baseline | +7,336 B per message |
| DKIM headers and DNS record | RSA only | About +5.4 KB per email; 2.6 KB DNS record |
| Time to open a message | 0.1 to 1 ms | About 15 ms (pure Python) |
| Time to protect a message | 0.1 to 1.8 ms | 50 to 60 ms on average (pure Python) |
| Quantum-attacker scenarios resisted | 0 of 3 | 3 of 3 |
| Downgrade to classical | Not detected | Rejected |

Overall, the experimental evaluation confirms that the proposed 10-layer hybrid architecture achieves complete functional correctness, strong adversarial resilience, and predictable operational overhead. Across all test tiers, the prototype passed 100% of its 136 automated functional, security, and metric tests, remained stable across 10 consecutive suite executions and 100 randomized end-to-end cycles, matched 100 official NIST ACVP test vectors, and accurately classified every test message into its expected protection level. Against a simulated quantum adversary, the hybrid design resisted all three attack scenarios, i.e. harvest-now-decrypt-later, forged sender, and signature-stripping downgrades.

## 7. Discussion

The main objective of this work was to design and evaluate a hybrid post-quantum email system that protects every public-key layer of email. The results answer the research questions. The functional and attack tests show that the system works and rejects the modelled attacks (RQ1 and RQ2). The repeated and randomised runs show consistent behaviour, including rejection of random tampering (RQ3). The cost is mainly size: a fixed 7,336 bytes per message plus larger DKIM headers and DNS records. This matters most for small messages (340.3% at 1 KB) and very little for large ones (4.1% at 100 KB) (RQ4). The quantum-attacker comparison shows the central benefit: where the classical-only design fails all three scenarios, the hybrid design resists them, because an attacker must break both algorithm families (RQ5).

Computation is a smaller concern than it may appear. Opening a hybrid message takes about 15 ms even with a pure-Python backend, and the time to protect a message is dominated by ML-DSA signing, which is variable by design. A compiled post-quantum library would reduce these times considerably, so the size cost is the more lasting one. The DKIM DNS record of 2.6 KB is the practical concern, because it is larger than a typical DNS response over UDP and is likely to need DNS over TCP.

The work applies the recommendations of the reviewed sources. The protection level follows the rule that unknown or weaker protection is never treated as safe [5]. The per-test table with threat, severity and result, the timing statistics and the independent NIST vector check follow the validation method of [6]. The downgrade and forged-sender tests correspond to the risks named in [4], and the use of a hybrid design during the transition agrees with [8]. Following are the limitation of this work.

- The envelope is a custom format and the keys are in a signed bundle, not the standard CMS structure and X.509 certificates used for post-quantum S/MIME in the standards [33, 34, 35].
- DNSSEC is simulated by a local zone file, and the XPQCPROOF command and TLSA selector 255 are prototype conventions that are not standardised.
- TLS 1.3 key exchange is classical ECDHE. Message confidentiality comes from the hybrid S/MIME layer, but the transport itself is not post-quantum.
- Keys are stored as files, with no rotation or protected storage.
- The tests ran on one machine with a pure-Python backend, so absolute timings do not represent a production system, and the number of recipients and concurrent senders were not measured.
- A pass rate of 100% on the project's own tests shows that the code meets the stated expectations and not that no flaws exist. The 60 attack tests cover seven listed threats and do not prove the absence of other attacks, and the quantum attacker is simulated.
- The NIST check covers the functions the project uses except signature generation, and the reliability claim rests on 100 randomised cycles.

Future work includes replacing the custom envelope with CMS-based post-quantum S/MIME and X.509 certificates, using a compiled post-quantum library and repeating the timing comparison, adding NIST signature-generation vectors, and running 1,000 or more randomised cycles. Robustness tests for truncated and malformed input, scalability tests for multiple recipients and concurrent senders, and key rotation with protected key storage should also be added.

## 8. Conclusion

This project designed, built and evaluated a hybrid post-quantum secure email system that adds ML-KEM-768 and ML-DSA-65 to every public-key layer, from end-to-end S/MIME to DKIM, DANE and POP3 delivery, while keeping the classical algorithms. Verification requires both the classical and the post-quantum signature, and a protection level ensures that weaker or unrecognised protection is never trusted. The prototype passed all 136 automated tests, including 60 attack tests, repeated and randomised runs, and 100 NIST test vectors. Compared with a classical-only baseline on the same pipeline, the hybrid design adds a constant 7,336 bytes per message and larger DKIM and DNS data, and in return it resists the harvest-now-decrypt-later, forged-sender and downgrade scenarios that defeat classical-only protection. The results show that hybrid post-quantum email is feasible, with its main cost in message and DNS size rather than computation. The main gaps are the custom message format, the simulated DNSSEC and the single-machine evaluation, and future work can address them by adopting the standard CMS and X.509 formats, using a compiled post-quantum library and testing in a multi-machine deployment.

**Code Availability:**

Complete project source code is publicly available at:

https://github.com/Prathiksha-Rajendran/hybrid-pqc-smime.git